\documentclass[3p,times,procedia]{elsarticle}
\usepackage{nupha_ecrc}

\volume{00}

\firstpage{1}

\journalname{Nuclear Physics A}

\runauth{}

\jid{nupha}

\jnltitlelogo{Nuclear Physics A}

\usepackage{amssymb}
\usepackage{array}
\usepackage{graphicx}
 \usepackage{multicol}

\usepackage[figuresright]{rotating}

\begin{document}

\begin{frontmatter}



\dochead{XXVIIth International Conference on Ultrarelativistic Nucleus-Nucleus Collisions\\ (Quark Matter 2018)}

\title{Z+jet productions in heavy-ion collisions}


\author[add1]{Shan-Liang Zhang\footnote{zhangshanl@mails.ccnu.edu.cn}}
\author[add1]{Tan Luo}
\author[add1,add2]{Xin-Nian Wang}
\author[add1]{Ben-Wei Zhang\footnote{bwzhang@mail.ccnu.edu.cn}}
\address[add1]{Key Laboratory of Quark \& Lepton Physics (MOE) and Institute of Particle Physics, Central China Normal University, Wuhan 430079, China}
\address[add2]{
Nuclear Science Division Mailstop 70R0319, Lawrence Berkeley National Laboratory, Berkeley, CA 94740}

\begin{abstract}
We report a systematic study of Z+jet correlation in Pb+Pb collisions at the LHC by combining the next-leading-order matrix elements calculations with matched parton shower in Sherpa for the initial Z+jet production, and Linear Boltzmann transport Model for jet propagation in the expanding quark-gluon-plasma. Our numerical results can well explain CMS measurements  on Z+jet correlation in Pb+Pb collisions: the shift of $p_T$ imbalance $x_{jZ}=p_T^{jet}/p_T^Z$ and their mean values, the suppression of the average number of jet partners per Z boson $R_{jZ}$, as well as  the modification of azimuthal angle correlations $\Delta \phi_{jZ}$.  We also demonstrate that high-order corrections play a significant role in the understanding of Z+jet correlations at high energies.

\end{abstract}


\end{frontmatter}


\section{Introduction}
\label{}
 Z boson tagged jet has long been regarded as a golden channel to study jet quenching\cite{Kartvelishvili:1995fr}. Though a fast parton from hard scattering lose energy due to  interactions with the hot medium when traveling through the quark-gluon plasma (QGP)~\cite{Gyulassy:2003mc}, Z boson will not participate in the strong-interactions directly and its mean free path is much longer than the size of the medium, escaping the QGP unscathed. Besides, Z boson have no contributions from fragmentation and decay because of its large mass ($M_Z=91.18$ GeV) as compared to photon, which may give some advantage for Z+jet  over photon+jet~\cite{Vitev:2008vk, Dai:2012am, Wang:2013cia, Neufeld:2010fj}.

The transverse momentum imbalance $x_{jZ}$, average number of tagged jet per Z boson $R_{jZ}$, azimuthal correlations $\Delta \phi_{jZ}$ in p+p and Pb+Pb collisions at 5.02 TeV have been measured by CMS experiment~\cite{Sirunyan:2017jic}. It is noted when computing the Z+jet azimuthal angle correlations, the next-leading-order (NLO) calculations suffer divergency at the region $\Delta \phi_{jZ}\sim \pi$, where a resummation of large logarithm may be required. Furthermore, leading-order (LO) matched parton shower (PS) calculations underestimate the azimuthal angle correlation at small angle region where wide angle radiation relative to the opposite direction of Z boson from higher order corrections are needed.  Motivated by this, we present in this talk a state-of-art calculations of Z+jet~\cite{Zhang:2018urd}, with p+p baseline compuated by the NLO+PS Monte Carlo~\cite{Gleisberg:2008ta}, and the Linear Boltzmann Transport (LBT) model~\cite{Wang:2013cia,He:2015pra} for jet propagation in heavy-ion collisions.

\section{Model setup for Z+jet in heavy-ion collisions}

We use Sherpa~\cite{Gleisberg:2008ta} to generate initial Z+jet events in p+p collisions at $\sqrt {s}=5.02 $ TeV.
Sherpa is a  Monte Corlo event generator, with which, low jet multiplicities can be calculated at NLO matched to the resummation of the PS~\cite{Hoche:2010kg, Hoeche:2012yf}.

\begin{figure}
  \centering
       \includegraphics[width=2.58in,height=2.16in]{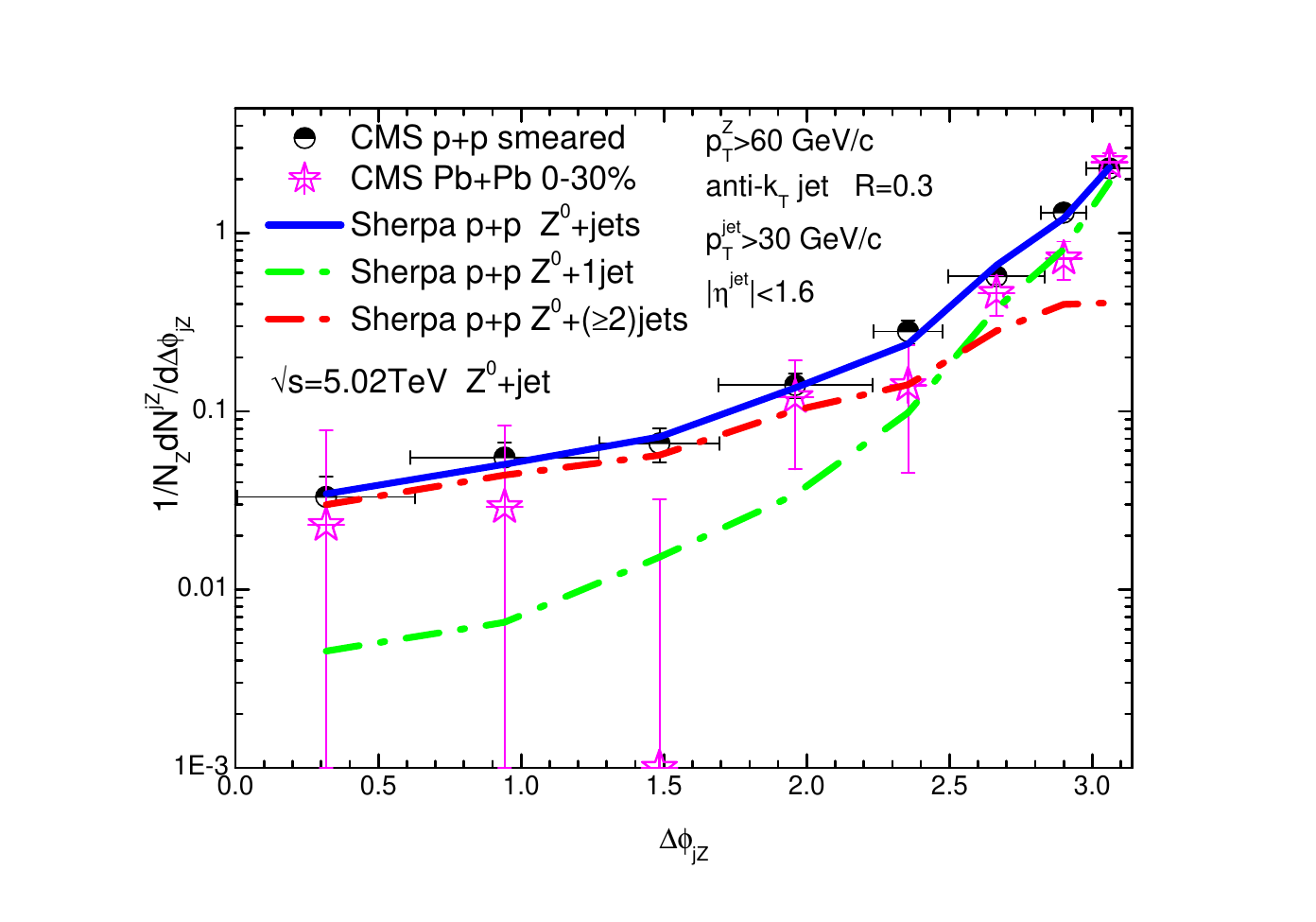}
       \includegraphics[scale=0.31]{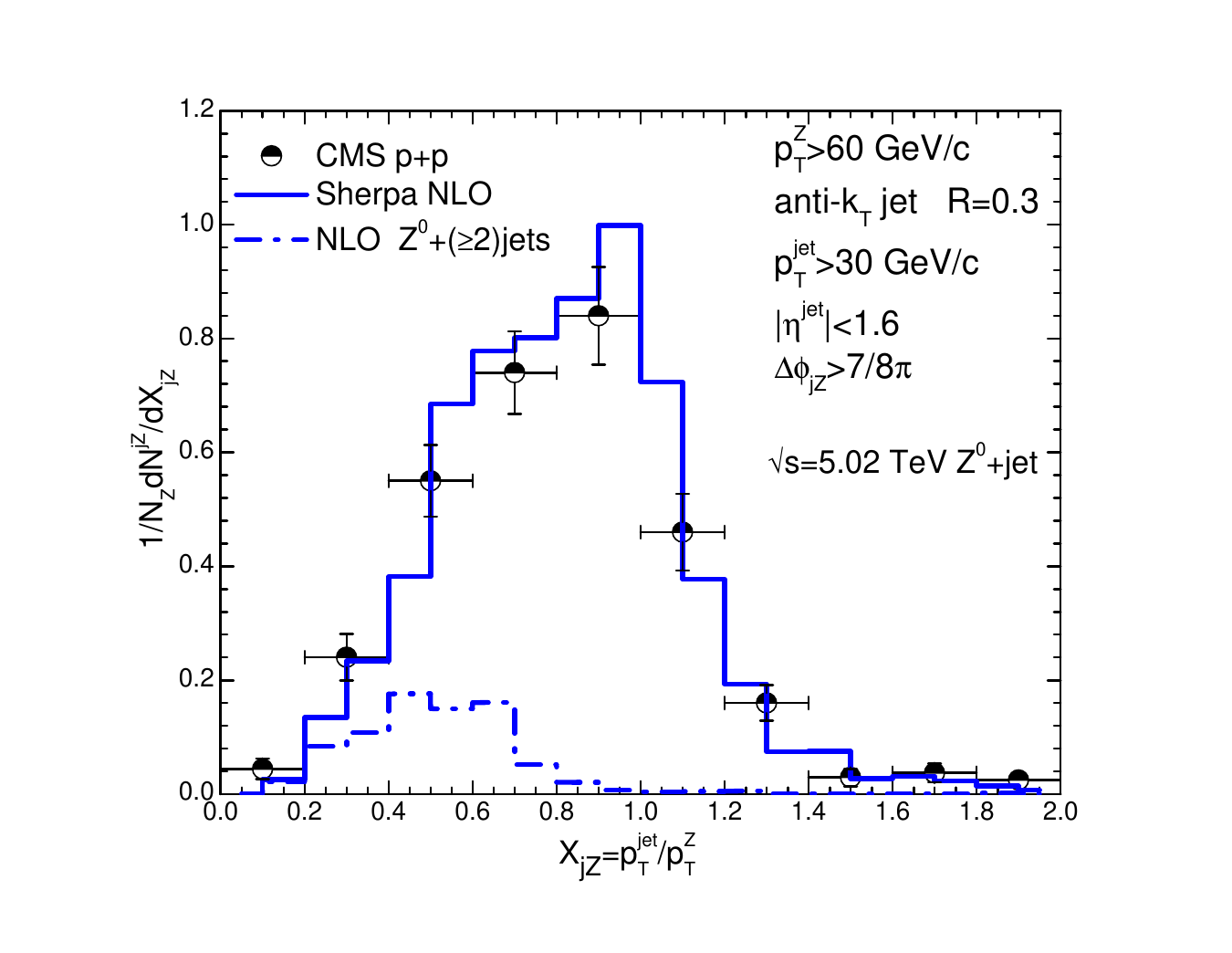}
             \vspace{-10pt}
  \caption{(Color online) The azimuthal angle correlations $\Delta\phi_{jZ}$ (left) and transverse momentum imablance $x_{jZ}$ (right) of Z+jet are calculated at 5.02 TeV by Sherpa  in p+p collisions and compared to  CMS experimental data.  }\label{zjetpp}
\end{figure}

 Fig.~\ref{zjetpp} illustrates the Z+jet correlations by Sherpa and the comparison with the CMS data in p+p collisions at 5.02 TeV.
NLO +PS calculations show excellent agreements with experiment in all kinetic ranges.  Contributions from  Z + only one jet  and Z plus more then one jet to the azimuthal angle  correlations are also plotted.  It indicates that Z + only one jet contributions dominates the large angle region,  while Z+ multi-jet dominates the small angle region. Numerical simulations of transverse  momentum imbalance by Sherpa describe the experimental data nicely.

In our model jet  propagation and induced medium response in hot QGP are simulated with the Linear Boltzmann Transport (LBT) model~\cite{Wang:2013cia,He:2015pra} which includes both elastic and inelastic scattering and follows the propagation of not only  jet shower partons and radiated gluons, but also the thermal recoil partons, in particular,  back-reaction of the Bolzmann  transport to keep energy-momentum conservation.
Elastic scattering is introduced by  $2 \rightarrow 2$ scattering matrix element $|M_{ab\rightarrow cd}|$ ,
and the inelastic scattering  is described by High-twist formalism for induced gluon radiation as
~\cite{Guo:2000nz, Zhang:2003wk, Schafer:2007xh},
\begin{equation}
\centering
\frac{dN_g}{dxdk_\perp^2 dt}=\frac{2\alpha_sC_A P(x)k_\perp^4}{\pi(k_\perp^2+x^2M^2)^4 }\hat{q}\sin^2\left(\frac{t-t_i}{2\tau_f}\right) \,\, .
\end{equation}
Here $x$ and $k_\perp$ are the energy fraction and transverse momentum of the gluon respectively,
 $P(x)$ and $\hat{q}$ are splitting functions and transport coefficient, $\tau_f=2Ex(1-x)/(k_\perp^2+x^2M^2) $ is the formation time of the radiated gluon.

\begin{figure}
  \centering
   \includegraphics[scale=0.22]{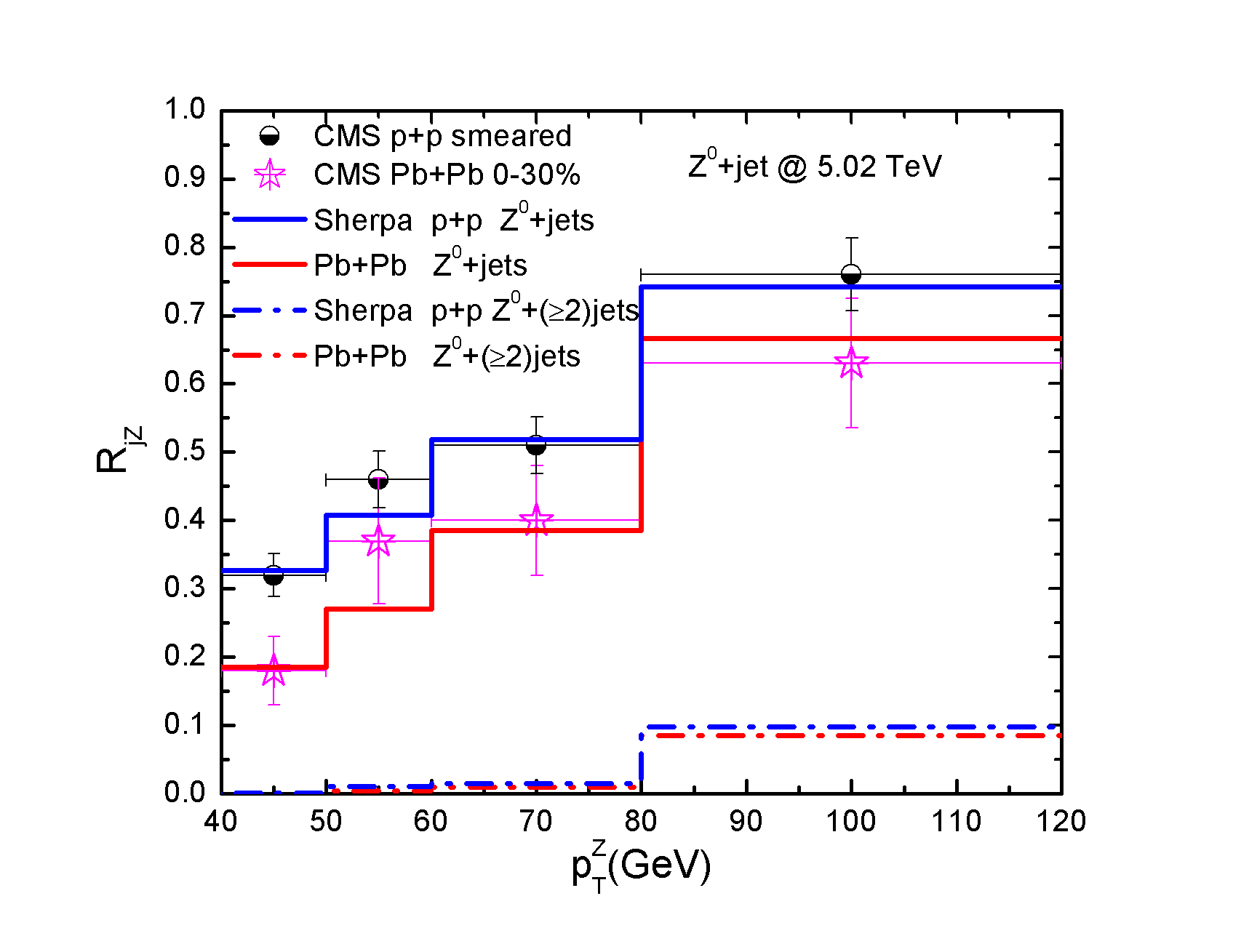}
   \includegraphics[scale=0.22]{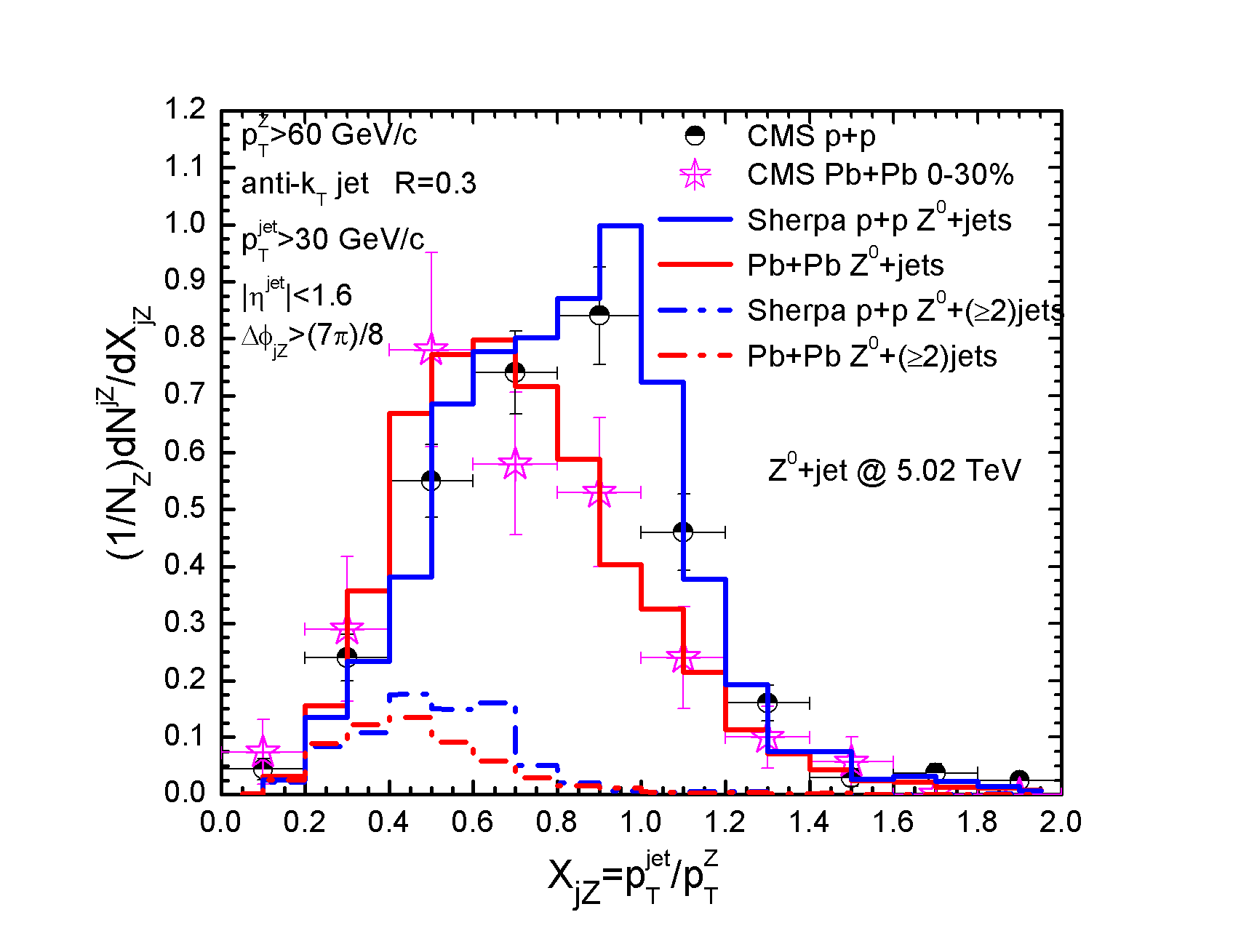}
         \vspace{-10pt}
  \caption{(Color online) (Left) Distributions of  $R_{jZ}$ (left) and $x_{jZ}=p_T^{jet}/p_T^Z$ (right)
  in central Pb+Pb collisions and p+p collisions at $\sqrt {s}=5.02 $ TeV.
  }\label{rjz}
\end{figure}

 \section{Results}
 In our numerical calculations, the Z boson and jets are selected according to the kinematic cut adopted by CMS~\cite{Sirunyan:2017jic}.  All the final partons, jet shower, radiated and medium recoiled partons are used to reconstruct jet. The underlying event background energy is subtracted event-by-event for Pb+Pb collisions following the procedure adopted by CMS~\cite{Kodolova:2007hd}, while no subtraction is applied in p+p collisions. And cold nuclear matter effects are found rather small for Z+jet in Pb+Pb at the LHC~\cite{Ru:2016wfx}.

Fig.~\ref{rjz} (left) shows $p_T^Z$ distribution of  average number of jet parters per Z boson $R_{jz}$. $R_{jZ}$ is overall suppressed in Pb+Pb, because large fraction of jets lose energy and then shift their final transverse momenta below the threshold $p^{jet}_T=$ 30 GeV.  We choose $\alpha_s=0.2$  to best describe experimental data of $R_{jZ}$  in Pb+Pb.   To select the most back-to-back Z+jet pairs, $\Delta \phi_{jZ}\ge \frac{7\pi}{8}$ is imposed, where the contribution from Z plus one jet processes dominates. The contribution of multijets to $R_{jz}$ is small in both p+p and Pb+Pb. For high energy Z boson with $p_T^Z> $ 80 GeV,  the processes from multi-jets give $\sim 15\%$ contribution.

Fig.~\ref{rjz} (right) plots the asymmetry distributions $x_{jZ}=p_T^{jet}/p_T^Z$.  Compared to p+p collisions, the asymmetry distribution in $x_{jZ}$ is broadened and shifted to the left in Pb+Pb collisions, due to jet energy loss in the medium while the transverse momentum of Z boson remains the same. Multi-jets processes give
$\sim 50 \%$  contributions when $x_{jZ}<0.5$ ,  where the energy of multi-jets can hardly exceed half of the energy of Z boson because of the  kinematic constraint  $\Delta \phi_{jZ}\ge \frac{7\pi}{8}$.

\begin{figure}
  \centering
   \includegraphics[scale=0.23]{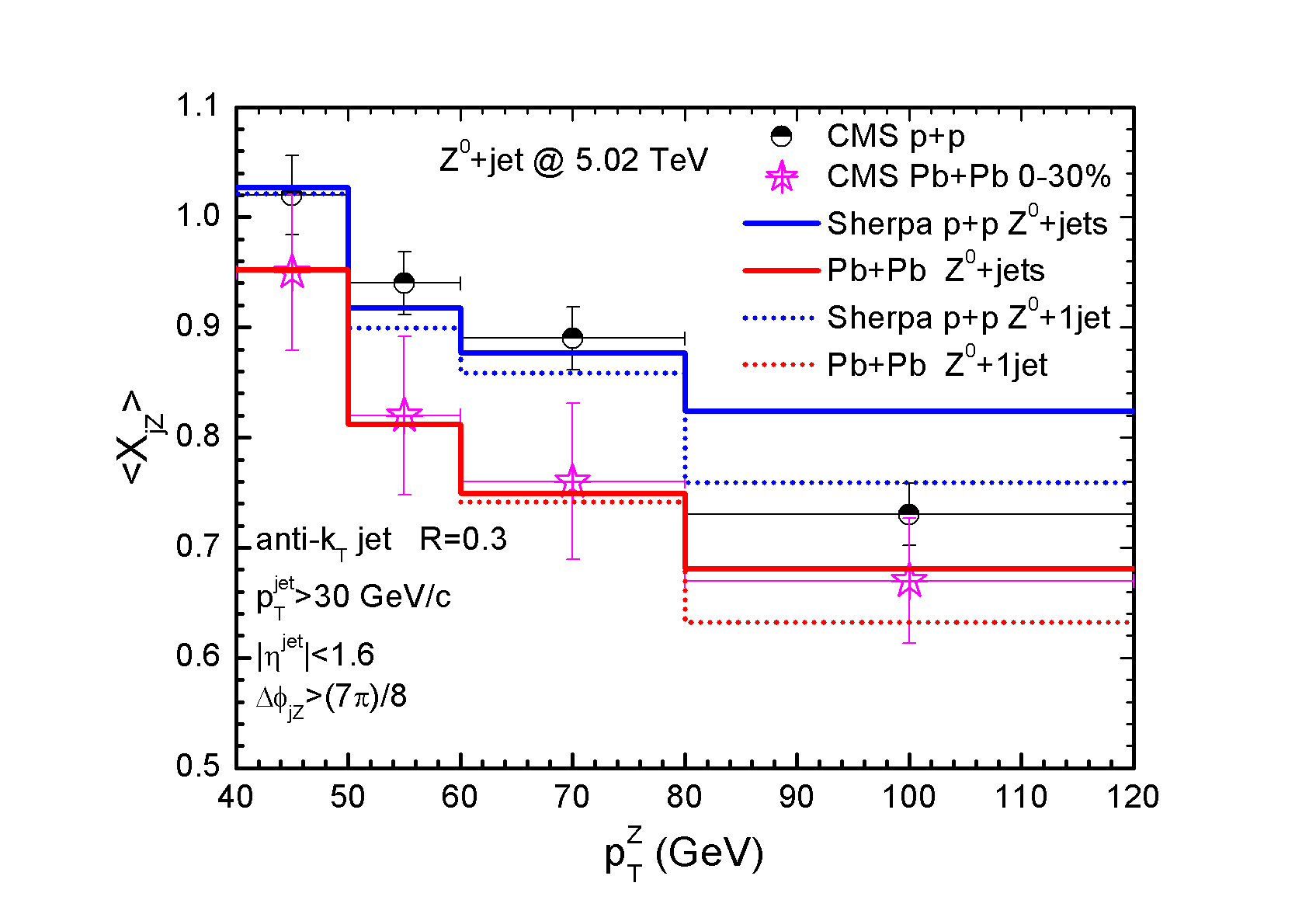}
   \includegraphics[scale=0.23]{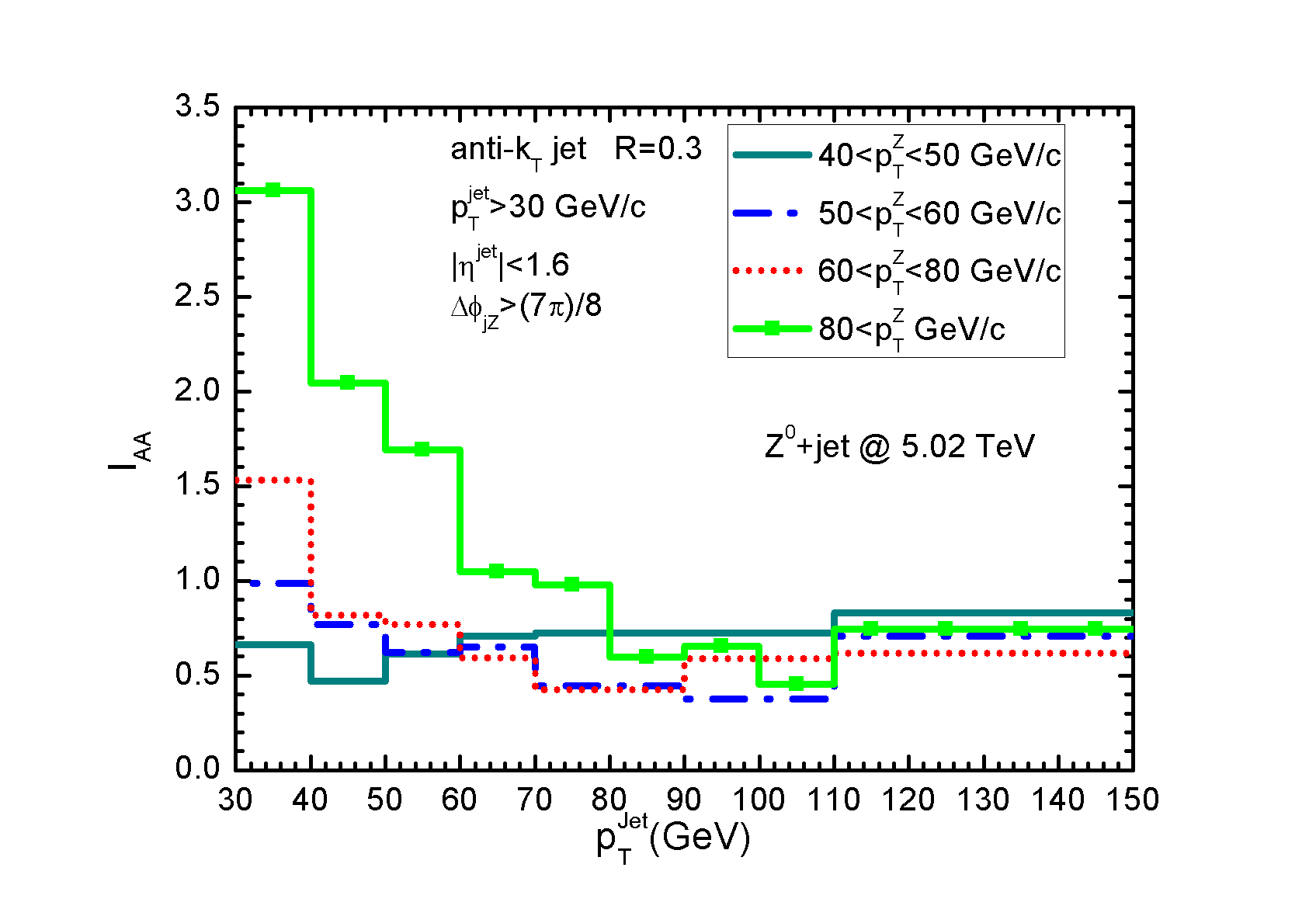}
      \vspace{-10pt}
  \caption{(Color online) (Left) Mean value of momentum imbalance $\langle x_{jZ}\rangle$ of Z+jet in central Pb+Pb collisions (red) and p+p collisions (blue) at $\sqrt{s}=5.02$ TeV.
   (Right) $I_{AA}$ as a function of $p_T^{jet}$ with different $p_T^Z$ bins.  }\label{xjz}
\end{figure}

In order to quantify the relative shift between p+p and Pb+Pb collisions, the mean value of momentum imbalance in different $p^Z_T$ bins are also calculated for completeness and shown in Fig.~\ref{xjz} (left). The mean value in Pb+Pb collisions is much smaller than that in p+p collisions.   For $p_T^Z >60$~GeV, the mean value is reduced by almost  $15\%$. We have also calculated  $\Delta \langle x_{jZ}\rangle=\langle x_{jZ}\rangle_{p+p}-\langle x_{jZ}\rangle_{Pb+Pb}$ in Table.~\ref{table:xjz}, which is consist with CMS data within uncertainty.  In Fig.~\ref{xjz} (right), we show the nuclear modification factor $I_{AA}=(dN^{Pb+Pb}/dp_T^{jet})/(dN^{p+p}/dp_T^{jet})$ of Z boson tagged jets in four $p^Z_T$ bins . We find that $I_{AA}$  is sensitive to kinematic cut. The strongest suppression is observed at $p_T^{jet} \simeq p^Z_T$, an enhancement in $p_T^{jet} < p^Z_T$ region,  and then a suppression in  $p_T^{jet} > p^Z_T$ region due to the steeper falling cross section in the kinematic cut region.

\begin{table}
\begin{center}
\begin{tabular}{|p{2.0cm}<{\centering}|p{2.0cm}<{\centering}|p{2.0cm}<{\centering}|p{2.0cm}<{\centering}|p{2.0cm}<{\centering}|}
  \hline

   $p_T^Z ({\rm GeV})$& 40-50 & 50-60 & 60-80 & $>$ 80 \\
      \hline
  ${\rm CMS \,\,\,\, data}$ & $0.07\pm0.106 $&0.12$\pm$0.148 & 0.13$\pm$0.158 &  0.06$\pm$0.088 \\
  \hline
  $\Delta \langle x_{jZ}\rangle$ &0.075&0.106 &0.128 &0.143 \\
  \hline
\end{tabular}
\label{table:xjz}
\caption{Relative shift of $\langle x_{jZ}\rangle$ between p+p collisions and Pb+Pb collisions at 5.02 TeV and the comparison with CMS data. }
\end{center}
\end{table}

\begin{figure}[tpb]
 \begin{center} \centering
   \includegraphics[scale=0.15]{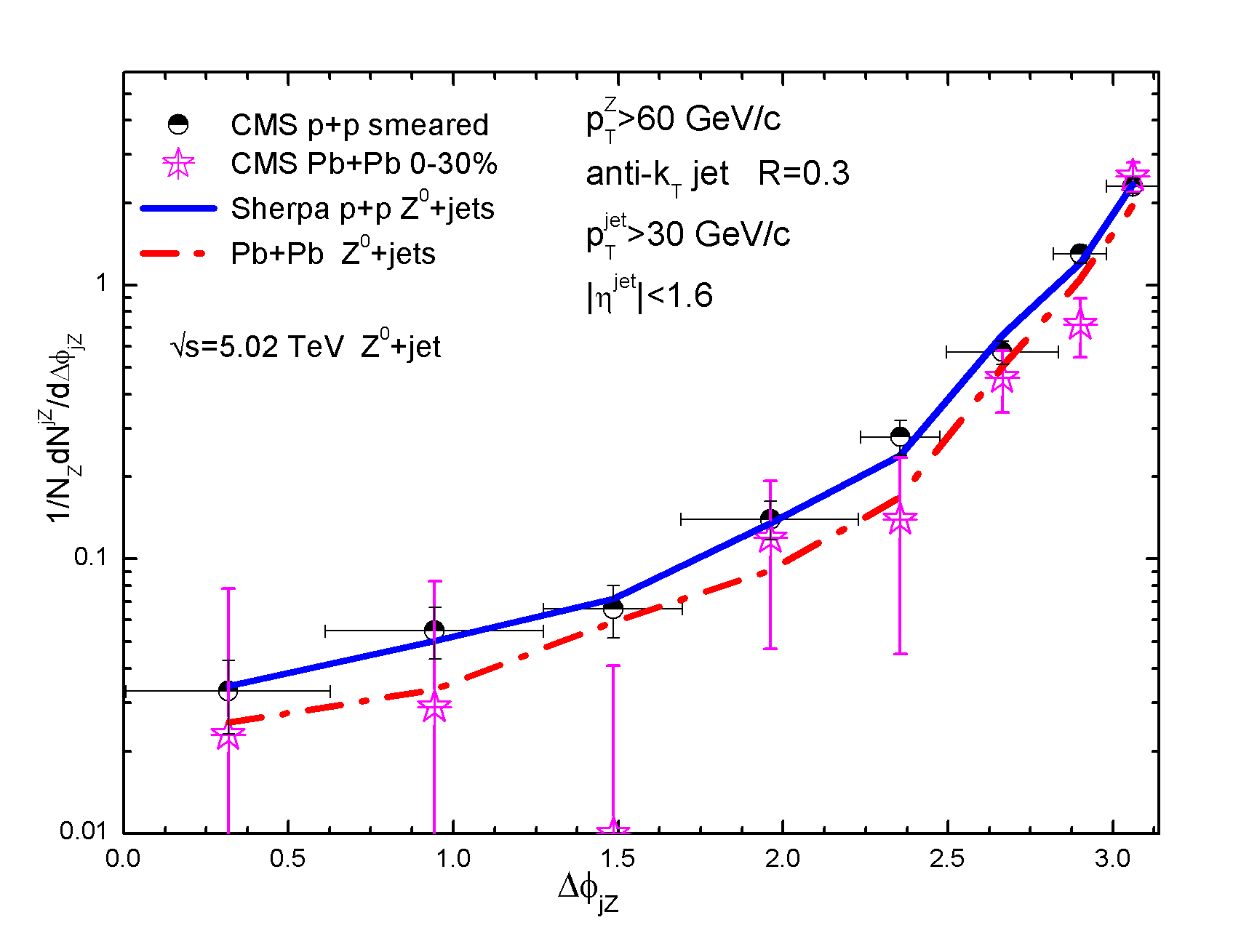}
   \includegraphics[scale=0.15]{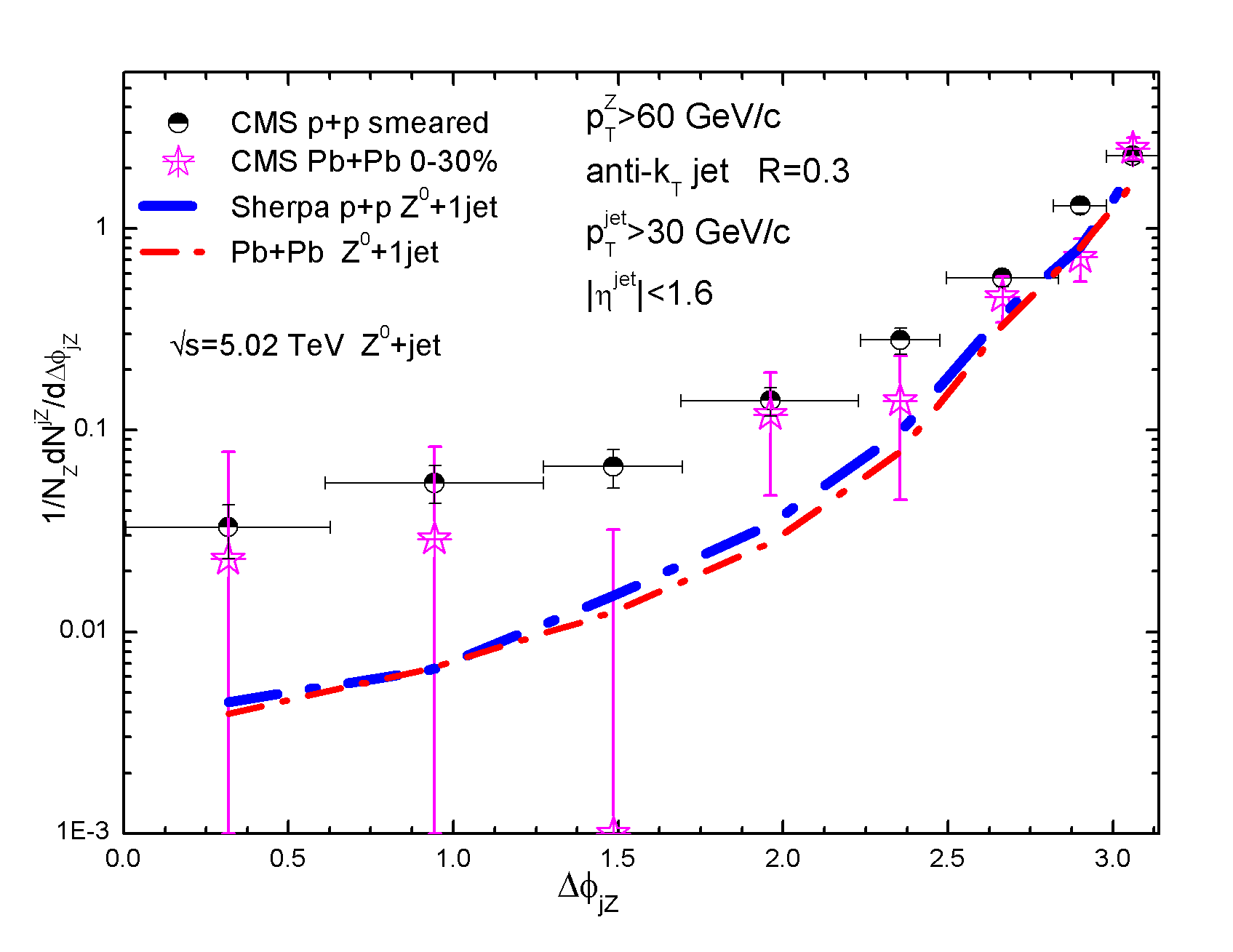}
   \includegraphics[scale=0.15]{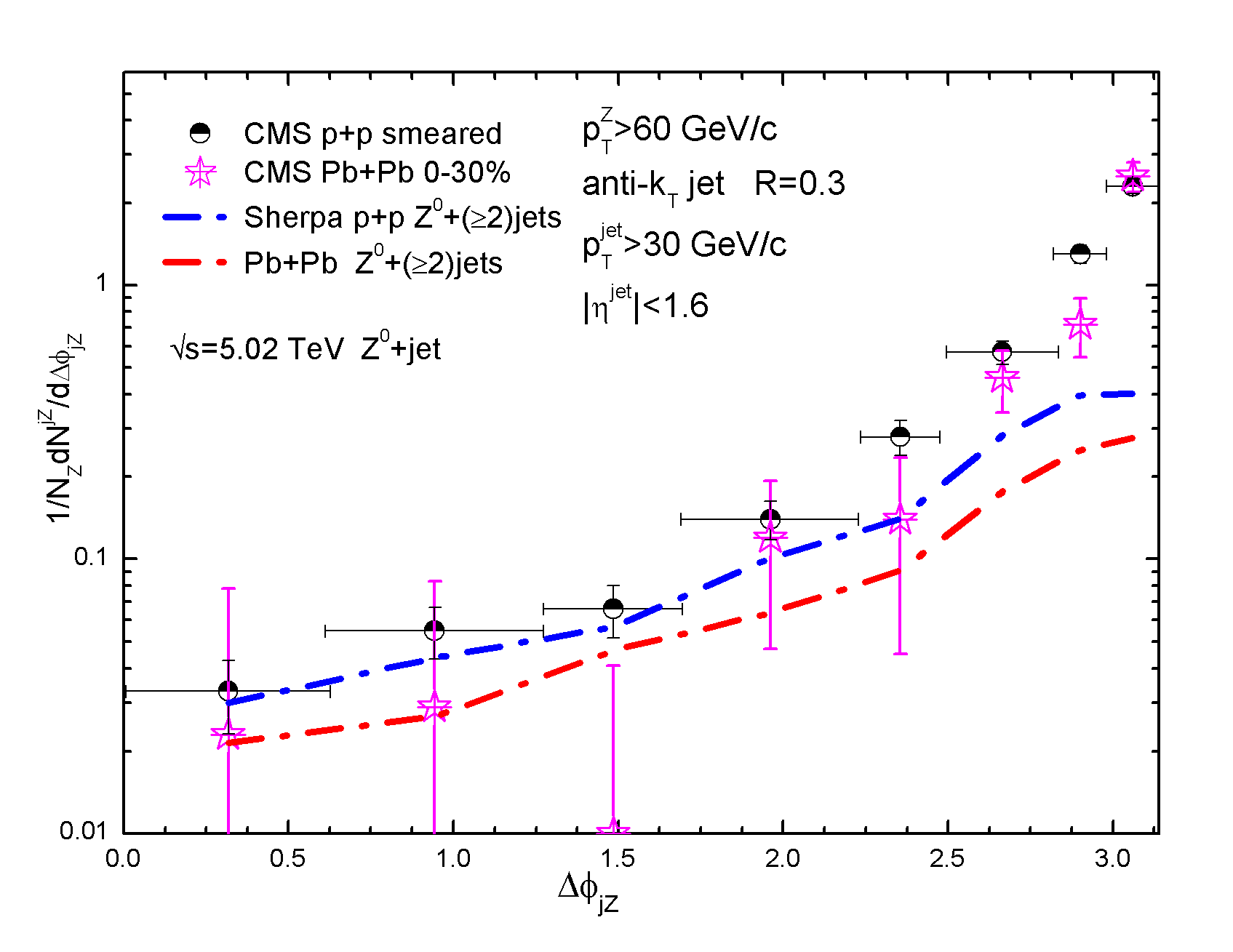}
 \end{center}
   \vspace{-10pt}
  \caption{(Color online) Z+jet azimuthal angle correlations $\Delta\phi_{jZ}=|\phi_{jet}-\phi_Z|$ (left), and the contributions  from Z plus only one jet (middle) and Z plus more than one jet (right) both in central Pb+Pb collisions and p+p collisions at $\sqrt{s}=5.02$ TeV.   }\label{phi}
\end{figure}

The numerical results of Z+jet azimuthal angle correlations $\Delta\phi_{jZ}=|\phi_{jet}-\phi_Z|$   in p+p and Pb+Pb are shown in Fig.~\ref{phi} (left).  It is moderately suppressed in Pb+Pb collisions. To illustrate the suppression mechanism, we also plot  separated contributions from Z+1jet and Z production associated with more than one jets in  both p+p and Pb+Pb collisions. In Fig.~\ref{phi} (middle), we present the contribution from Z + 1jet. We see  there is no significant difference for Z+1jet processes between p+p and Pb+Pb collisions.
In these processes, the transverse momentum of the Z boson is balanced by only one back-to-back jet and the azimuthal angle correlations  are focused mainly on  $\Delta\phi_{jZ}\simeq\pi$ region.  These processes mainly come from  the LO contribution and the  azimuthal angle decorrelation  from which is dominated by soft/collinear radiation. The transverse momentum broadening of jets due to jet-medium interaction is negligible at such high energy scale. Fig.~\ref{phi} (right)  illustrates the results of Z+ multijet, which is considerably suppressed in Pb+Pb collisions.
Compared to Z+1jet,  azimuthal angle correlation from Z+ multijet is broadened and becomes flat.  The main  contribution of  Z+ multijet comes from NLO ME processes. The transverse momentum balance of Z boson and the jet is broken in these processes with additional emissions of hard partons. The initial energy of the tagged jet is smaller compared to Z boson, and it can easily fall below $p^{jet}_T$ = 30 GeV threshold in Pb+Pb due to jet energy loss effect. It is the suppression of multijets that mainly leads to the modification of Z+jet azimuthal angle correlation.

This work has been supported by NSFC of China with Project Nos. 11435004 and 11521064, MOST of China under 2014CB845404, NSF
 under grant No. ACI-1550228 and U.S. DOE under Contract No. DE-AC02-05CH11231.





\bibliographystyle{elsarticle-num}



\end{document}